# Context-aware QR-codes

Dmitry Namiot
Lomonosov Moscow State University
Faculty of Computational Math and Cybernetics
Moscow, Russia
e-mail: dnamiot@gmail.com

Manfred Sneps-Sneppe
Central Telecommunication Research Institute (ZNIIS)
M2M Competence Centre
Moscow, Russia
e-mail: manfreds.sneps@gmail.com

Oleg Skokov
Central Telecommunication Research Institute (ZNIIS)
M2M Competence Centre
Moscow, Russia
e-mail: oskokov@zniis.ru

*Abstract*— **This paper describes a new model for presenting local information based on the network proximity. We present a novelty mobile mashup which combines Wi-Fi proximity measurements with QR-codes. Our mobile mashup automatically adds context information the content presented by QR-codes. It simplifies the deployment schemes and allows to use unified presentation for all data points, for example. This paper describes how to combine QR-codes and network proximity information.**

## I. INTRODUCTION

QR Codes or two dimensional (2D) barcodes are used to encode and decode data at a rapid rate. Using camera phones and appropriate applications to read 2D barcodes for various purposes is currently a widely- used approach in practical applications [1]. Mostly, QR codes provide some static information (link, text, SMS dialog or vCard info).

QR codes storing addresses and Uniform Resource Locators (URLs) may appear in magazines, on signs, on buses, on business cards, or on almost any object about which users might need information. Users with a camera phone equipped with the correct reader application can scan the image of the QR code to display text, contact information, connect to a wireless network, or open a web page in the telephone's browser. This act of linking from physical world objects is termed hardlinking or object hyperlinking. QR codes also may be linked to a location to track where a code has been scanned. Either the application that scans the QR code retrieves the geo information by using GPS and cell tower triangulation (aGPS) or the URL encoded in the QR code itself is associated with a location [2]. Obtaining geo information at the moment of QR code scanning is an example for dynamic data support in QR codes deployment models. This paper proposes a particular use case which merges a public QR Code and private information, in order to provide data related to a particular context. This model lets use unified QR code installtion across all places indoor. At the decoding stage this public info will be translated into context aware presentation.

There are several papers devoted to context-aware QR codes. As per classical Dey's definition, a context-aware system is a computing system using context to provide relevant information and/or services to the user, where relevancy depends on the user's tasks [3]. One of the main goals of ubiquitous computing is to provide relevant information, at the right time and place and under the right form. Context-aware systems should help filter information. QR codes provide on-place information for mobile users. By this reason, they are a perfect example for ubiquitous computing models.

J. Rouillard in [1] combiles context-aware QR codes from two parts: "traditional" QR code info and XML-based description for messages that should be provided to the reader. It is presented on the Figure 1.

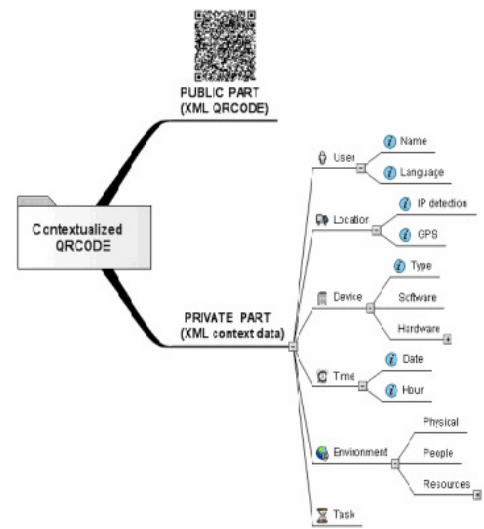

Figure 1. Contextualized QR code [1]

The private part can be one or more information among the subsequent: user's profile, current task, device used, location, time and environment of the interaction. The machine decodes the QR Code and merges it with private data obtained during the interaction. Then, the XML resulting file is sent to a web service (created in author's laboratory) that computes the code and returns personalized messages. Conceptually this approach is close to context-aware messaging system [4], but in this case author did not disclose the way context info could be obtained. We can note also, that except the location, it is not a fully context info.

F. Lyardet, et. al. [5] use QR codes for context-aware navigation. When using QR codes as a location source, the physical location of the QR code is encoded into the URL. The context server is then used to map location codes to symbolic locations.

There are models of applications, where the concept of location can be replaced by that of proximity. At the first hand, this applies to use cases where the detection for exact location is difficult, even impossible or not economically viable [6]. Very often, this change is related to privacy. For example, a privacy-aware proximity detection service determines if two mobile users are close to each other without requiring them to disclose their exact locations [7].

Metric measurements for privacy can be replaced with some approximation by wireless proximity (network proximity). For this paper network proximity definition is very intuitive. It is a measure of how mobile nodes are close or far away from the elements of network infrastructure. There are several systems that can use network proximity as a base for mobile services. At the first hand, we can mention here our own system SpotEx (Spot Expert) [8]. According to this model, any existing or even especially created Wi-Fi hot spot could be used as presence sensor that can trigger access for some user-generated information snippets.

The typical application in this area uses collected database of so called Wi-Fi "fingerprints". Each fingerprint desribes some place and contains MAC addresses and the received signal strengths (RSSI) of nearby access points. At the first hand, this database could be used for Wi-Fi based positioning. At the second, historical records for fingerprints let us discover user's behavioral patterns [9]. A classical approach to Wi-Fi fingerprinting [10] involves RSSI (signal strength). The basic principles are transparent. At a given point, a mobile application may hear ("see") different access points with certain signal strengths. This set of access points and their associated signal strengths represents a label ("fingerprint") that is unique to that position. The metric that could be used for comparing various fingerprints is k-nearest-neighbors in signal space. Figure 2 [10] shows response rate for a single AP as a function of distance from that AP. Response rate has been defined by authors as follows: the percentage of times that a given AP was heard in all of the Wi-Fi scans at a specific distance from that AP.

Problems associated with the collection of fingerprints, are fairly obvious. It is the price of the calibration process, the need for rework after the changes in the network and, most importantly, lack of support for dynamic networks. For example, most of the modern smart phones let users open Wi-Fi access point right on the phone. As soon as we link our data to dynamic access points, we can not use static bases of fingerprints [6].

More precisely, we can not use previously colelected fingerprints. But in the same time, we can still use various metrics for dinamically collected Wi-Fi info. In the classical Wi-Fi based positioning, on the positioning phase, a user's mobile device performs a scan of its environment. Location engine compares this scan with all of the fingerprints in the radio map to find the fingerprint that is the closest match to the positioning scan in terms of APs seen and their corresponding signal strengths [10]. The classical approach calculates the Euclidian distance in the signal space. As the next step it selects k records (fingerprints) in the radio map database with the smallest distance to the observed scan and computes the average of the latitude-longitude coordinates associated with those k fingerprints. On practice, the good accuracy could be reached with k = 4. In general, fingerprints based approach is based on the assumption that the Wi-Fi devices used for training and positioning measure signal strengths in the same way. Very often it is not so due to differences caused by manufacturing variations, antennas, orientation, humidity, etc.)

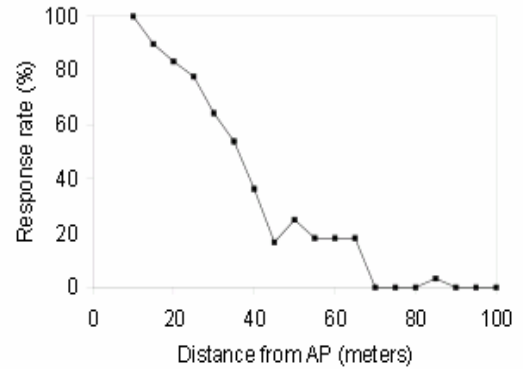

Figure 2. Response rate vs. distance

The modifications may include ranking systems. Instead of comparing absolute signal strengths, this method compares lists of access points sorted by signal strength. For example, if the positioning scan discovered (P1, P2, P3) with signal strengths (-30, -80, -40), then we replace this set of RSSI strengths by their relative ranking, that is (1, 3, 2). Like-wise, if the positioning scan discovered (P4, P5, P6) with signal strengths (-50, -20, -40), then we replace this set of RSSI strengths by their relative ranking, that is (3, 2, 1). Then we can compare the relative rankings using the Spearman rank-order correlation coefficient:

$$\frac{\sum_i (Ri - \overline{R})(Ri' - \overline{R'})}{\sqrt{\sum_i (Ri - \overline{R})^2} \sqrt{\sum_i (Ri' - \overline{R'})^2}}$$

Using the relative order of signal strengths in this way means that fingerprints will still match well in spite of differences in scale, offset, or any monotonically increasing function of signal strength separating the Wi-Fi devices [10].

Finally, we can list the following measurements that could be used for distance comparison: the visibility of AP, RSSI (signal strength) and AP's rank.

Staistical methods can use the information about the probability for the particular AP to be visible from this place. This information could be used for the positioning too. Thus, comparing the distances, we can use the fact of the presence of access points in the list of visible access points, signal strength and AP's rank.

Statistical methods may involve more data and see the information about the likelihood of a specific access point from a certain position. This information may also be used for position determination. For example, Statistical Learning Theory could be applied to the problem of determining the location of a wireless device by measuring the RSSI values from a set of access points. The typical example is the Support Vector Machine paradigm [11]. Methods based on conditional probability estimation require the knowledge of the signal propagation model. At the first hand, it could be presented in the form of an empirical distribution from repeated observations on each physical point in a given set. Obviously, it requires a large number of observations in order to have a precise distribution estimate at each sample point. Alternatively, we can select a suitable radio propagation model and estimate its parameters on the basis of empirical observations. Once the radio propagation model is determined, it can be used to calculate the conditional probability distribution

The Bayes law of conditional probability can be used to calculate the inverse dependency, in the form of a probability distribution of the physical coordinates depending on signal strength information.

In general, statistical methods even more closely associated with the accumulation of historical material (training phase). It is not a good case for the dynamic environments.

Our main idea is to give a software tool that can be used to determine the information provided to mobile users based on proximity of the mobile subscriber. Practical use for proximity in mobile services will be possible only if software developers can work with proximity data as comfortable as with the location info. Convenience for the development includes two aspects: the ease of presentation and ease of processing.

II. CUSTOMIZED QR-CODE READER

Our idea about context-oriented QR-code scanner is based on the modified version of open sourced scanner Zxing (Zebra Crossing) [12].

In general, 2D barcodes encode some text. But in many cases, that text can represent many different things. Text representing contact information, when recognized, could trigger a prompt to add the contact to an address book, for example. Compared with 1D codes, 2D codes can save a larger amount of data. In addition, an advanced error-correction method allows the QR Code to be read more reliably and at high er speeds than other codes [13]. Figure 3 describes the structure for 2D barcode.

QR Code is a 2D matrix code that conveys information not by the size and position of bars and spaces in a single (horizontal) dimension, but by the arrangement of its dark and light elements in columns and rows. Each dark or light module of a QR Code symbol represents a 0 or 1.

In the most often usage, 2D barcodes encode text that represents a URL, like "*http://some_domain.com/*". Technically, this is a special string of text since it is recognizable as a URL by readers. Therefore, it can be acted upon, and the reader can open the URL in a browser.

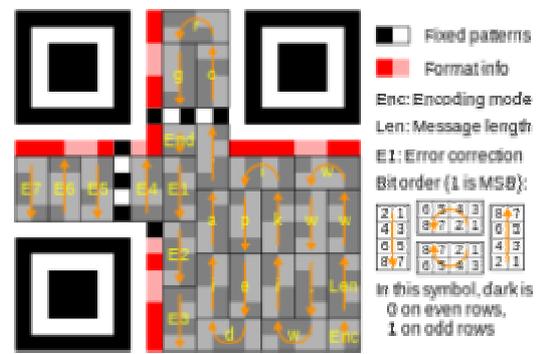

Figure 3. QR-code format

Our idea is to keep the basic recoginition as is, just to add some parameters on the final stage. In other words, customized QR-code scanner will replace encoded value *http://some_domain.com/* with *http://some_domain.com?list_of_parameters*. And this list of parameters will describe our context.

For example, let us see the QR-code deployment for some indoor retail application. Technically, we can prepare a separate QR-code for the each existing deparment, but it could be costly and difficulty to maintain. The feature that can automatically add context info to any encoded URL lets us use the same QR-code across all our installations. We can encode the generic URL (end point for our application) and the local information will be passed to our end point automatically.

On the top level the whole schema is transparent. QR-code scanner is a normal mobile application. This application as any another application has got access to the phone's features. More precisely, it can get that access (it depends on permissions allowed by the user). For example, mobile application can get access to location information. It means that encoded URL

*http://some_domain.com*

will be replaced by some like this, for example:

*http://some_domain.com?lat=latitude&lng=longitude*

where latitude and longitude are calculated values for user's location. For indoor deployment, the usage of GPS based positioning could be limited. Wi-Fi based positioning is limited by the need of collecting fingerprints a priori. By this reason we think about passing network proximity info to the target URL. It means that our customized QR-code scanner creates on the fly Wi-Fi fingerprint and adds this information to the query string for encoded URL.

For fingerprint info we can collect the following information:

SSID name
MAC-address
RSSI

This set describes one Wi-Fi network which is visible from the mobile phone. And our fingerprint is just a collection of such triples. So, in our case we will modify the encoded URL by this way:

*http://some_domain.com?fingerprint*

As an example illustrated the support for dynamic systems, we can present the following use case. Most of modern smart phones let users open Wi-Fi hotspots right on the phone (Figure 4).

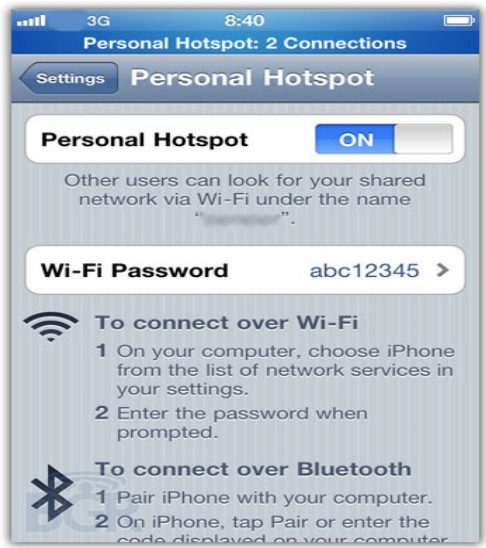

Figure 4. Mobile Hot Spot on the phonee

It is not necessary to open internet access there. Just having a hot spot is enough. Actually, we are using here our ideas for dynamic LBS presented in [6]. This hot spot can work as a presence trigger. And in the same time, QR-code itself could be demonstrated right from the mobile screen (Figure 5). Fingerprint for the scanned QR-code (for decoded URL) lets us easily identify the media (mobile phone) where QR-code has been presented (has been scanned from).

Suppose our encoded URL presents some CGI-script (dynamic web page). Context-related parameters could be easily analyzed in this script, because we pass them as any another GET parameter. In other words, we do not need to present any special Application Program Interface (API) for access to context data. Our API presented below juts simplifies data processing. But the basic access uses standard approach from HTTP protocol. It is about the convenience for the development. The presentation aspect is very simple for now.

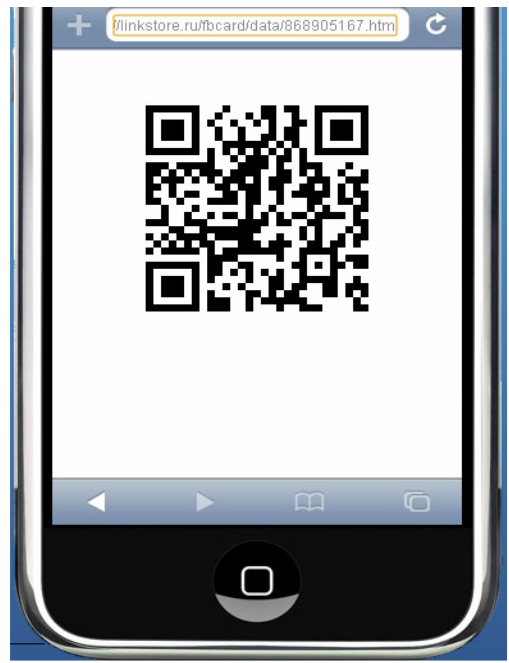

Figure 5. QR-code on the screen

Our description below will target the simplicity of processing context information.

### III. DYNAMIC FINGERPRINTS

In order to pass context data, we can choose one of two options. At the first hand, we can follow to the standard HTTP approach. We can add to the basic URL a series of GET parameters in the the following form:

*parameter_name = parameter_value*

Parameters will describe our fingerprint (MAC address, RSSI level for the each "visible" access point). From the developers point of view, access to the provided data in the server side script is absolutely transparent. For example, in Java servlets we can use the following calls from HttpServletRequest class:

*java.util.Enumeration<java.lang.String> getParameterNames()*

It returns an Enumeration of String objects containing the names of the parameters contained in this request.

*java.lang.String[] getParameterValues(java.lang.String name)*

It returns an array of String objects containing all of the values the given request parameter has.

So, processing context information is an ordinary task for web development domain. As a typical business related algorithm we can mention here SpotEx production system [14]. The typical use case could be for example idoor naviagtion systems based on QR-codes [15].

The second (and more prospect) approach could be based on the centralized storage for the dynamically collected fingerprints. It means that our customized QR-code reader will store collected fingerprints in some cloud based data store (e.g., relational database) and pass to the target URL just one ID for stored data. Server side script (e.g., Java Servlet) can use this ID for obtaining recorded context info. In other words, the source URL like

*http://domain.com*

will be replaced with

*http://domain.com?context_id=some_value*

On the one hand, this approach allows us to maintain the previous style of processing. Instead of the direct request for context parameters, we'll use the our datastore API to access the saved info. In this case, the ID will serve as a key.In our database we can keep the following info (per records):

*ID*
*Timestamp*
*MAC address*
*RSSI level*

So, we can use the following calls, for example:

*java.util.Enumeration<java.lang.String> getParameterNames(java.lang.String ID)*

*java.lang.String[] getParameterValues(java.lang.String ID, java.lang.String name)*

But the centralized storage can give a new quality. For example, it will analyze the history, behavior, identify new users. In the future, centralized repository will add behavioral API for users (how many times have scanned the address near-such points, how many times have scanned it at all, statistics scan time). Note, that the scanning will be still anonymous. QR-code users differ only in the MAC-address. This MAC-address is used only to reidentification. Hence, we can replace it with a hash code for saving privacy.

One example of the use of centrally stored data is the realization of a product system similar Spotex. Our server-side script can deploy some form of production based systems. For the each rule, conclusion (right part) describes some business action. Each condition, in the same time, could be presented as a logical expression with build-in functions. We can list the the following standard functions related to the context:

*COUNTER( )*
*FIRST ( )*

Function COUNTER (n) returns a number of scans for the current user. Its argument describes time interval. Possible values for time interval are:

0 - all time
1 – day
2 – week
3 – month

E.g., CONTER(2) returns the total number of scans for the current week.

Function FIRST(n) returns a boolean value true if the scan is first for the given time interval.

For example:

*IF COUNTER(3)>2 AND FIRST(2) THEN*
*{ deliver coupon info message }*

The typical use cases are proximity marketing and news delivery in Smart City projects for example.

IV. CONCLUSION

This article presents a new way for QR-code deployment. Our paper describes a new mashup combines 2D barcodes and network proximity data. The proposed model uses smart-phone as a proximity sensor and creates some stardard presentation for data. Described service can use existing as well as the especially created networks nodes as presence triggers for delivering and discovering a new content for mobile subscribers. Proposed services could be used for proximity marketing and delivering commercial information (deals, discounts, coupons) in malls, for indoor navigation, for access to hyper-local news data and for data discovery in Smart City projects.

REFERENCES

[1] Rouillard, J. (2008, July). Contextual QR codes. In Computing in the Global Information Technology, 2008. ICCGI'08. The Third International Multi-Conference on (pp. 50-55). IEEE.

[2] Bellucci, A., Ghiron, S. L., Aedo, I., & Malizia, A. (2008, May). Visual tag authoring: picture extraction via localized, collaborative tagging. In Proceedings of the working conference on Advanced visual interfaces (pp. 351-354). ACM.

[3] Dey. A., Providing architectural support for building context-aware applications. Ph.D thesis, College of computing, Georgia institute of technology, 2000

[4] Namiot, D., & Schneps-Schneppe, M. (2011, September). "About Location-aware Mobile Messages: Expert System Based on WiFi Spots". In Next Generation Mobile Applications, Services and

Technologies (NGMAST), 2011 5th International Conference on (pp. 48-53). IEEE. DOI: 10.1109/NGMAST.2011.19

[5] Lyardet, F., Szeto, D., & Aitenbichler, E. (2008). Context-aware indoor navigation. Ambient Intelligence, pp. 290-307

[6] D. Namiot and M. Sneps-Sneppe "Context-aware data discovery" Intelligence in Next Generation Networks (ICIN), 2012 16th International Conference on, 2012 pp. 134 – 141, DOI: 10.1109/ICIN.2012.6376016

[7] L.Šikšnys, J. Thomsen, S. Šaltenis, and M. Yiu Private and Flexible Proximity Detection in Mobile Social Networks, Mobile Data Management (MDM), 2010 Eleventh International Conference on, 2010, pp. 75 - 84

[8] D. Namiot "Context-Aware Browsing -- A Practical Approach", Next Generation Mobile Applications, Services and Technologies (NGMAST), 2012 6th International Conference on, pp. 18-23 DOI: 10.1109/NGMAST.2012.13

[9] J. Rekimoto, T. Miyaki, and T. Ishizawa. LifeTag: WiFi-based Continuous Location Logging for Life Pattern Analysis. in LOCA. 2007

[10] Y. Chen, Y. Chawathe, A. LaMarca, and J. Krumm. "Accuracy characterization for metropolitan-scale Wi-Fi localization". In ACM MobiSys, 2005.

[11] Brunato, Mauro, and Roberto Battiti. "Statistical learning theory for location fingerprinting in wireless LANs." Computer Networks 47.6 (2005): pp. 825-845.

[12] Zxing http://code.google.com/p/zxing/ Retrieved: Apr, 2013

[13] Denso QR-code essential http://www.denso-adc.com/pdf/qrcode Retrieved: Apr, 2013

[14] D. Namiot and M. Sneps-Sneppe "Proximity as a service". In Future Internet Communications (BCFIC), 2012 2nd Baltic Congress on, pp. 199-205. IEEE.

[15] Gao, J. Z., Prakash, L., & Jagatesan, R. (2007, July). Understanding 2d-barcode technology and applications in m-commerce-design and implementation of a 2d barcode processing solution. In Computer Software and Applications Conference, 2007. COMPSAC 2007. 31st Annual International (Vol. 2, pp. 49-56). IEEE